\documentclass[epj]{svjour}

\usepackage{graphics}
\usepackage{amsmath,amssymb,graphics}
\usepackage{amsmath,amssymb,graphics,epsfig,subfigure}

\onecolumn

\begin{document}

\title{Black hole solution and strong gravitational lensing in Eddington-inspired Born-Infeld gravity}

\author{Shao-Wen Wei \thanks{\email{weishw@lzu.edu.cn}} \and
        Ke Yang \thanks{\email{yangke09@lzu.edu.cn}}\and
        Yu-Xiao Liu \thanks{Corresponding author. \email{liuyx@lzu.edu.cn}}}
\institute{Institute of Theoretical Physics, Lanzhou University, Lanzhou 730000, People's Republic of China}

\abstract{
A new theory of gravity called Eddington-inspired Born-Infeld (EiBI) gravity was recently proposed by Ba\~{n}ados and Ferreira. This theory leads to some exciting new features, such as free of cosmological singularities. In this paper, we first obtain a charged EiBI black hole solution with a nonvanishing cosmological constant when the electromagnetic field is included in. Then based on it, we study the strong gravitational lensing by the asymptotic flat charged EiBI black hole. The strong deflection limit coefficients and observables are shown to closely depend on the additional coupling parameter $\kappa$ in the EiBI gravity. It is found that, compared with the corresponding charged black hole in general relativity, the positive coupling parameter $\kappa$ will shrink the black hole horizon and photon sphere. Moreover, the coupling parameter will decrease the angular position and relative magnitudes of the relativistic images, while increase the angular separation, which may shine new light on testing such gravity theory in near future by the astronomical instruments.}



\maketitle

\section{Introduction}

Einstein's theory of general relativity (GR) has achieved great success on interpreting many phenomenological and experimental results. However, one of its major riddles is the prediction of the appearance of the singularity, such as that in the beginning of Big Bang and at the center of black holes. In order to resolve the singularity problem, based on the classic work of Eddington \cite{Eddington}, and on the non-linear electrodynamics of Born and Infeld \cite{BornInfeld}, an alternative gravity theory, Eddington-inspired Born-Infeld (EiBI) theory, was recently proposed by Ban\~{a}dos and Ferreira \cite{BanadosFerreira}. Different from the metric Born-Infeld-Einstein theory \cite{Born-Infeld-Einstein} and the purely affine Eddinton theory \cite{Eddington}, EiBI gravity is a Palatini theory of gravity. For different Palatini theoreies see Refs. \cite{Sotiriou2010} and \cite{Makarenko2014} for Palatini $f(R)$ and Born-Infeld-$f(R)$ gravities, respectively. For recent reviews of modified gravities, see Refs. \cite{Nojiri2011PhysRept,Clifton2012,Capozziello:2011et}.

EiBI gravity has aroused much interest. It was found that EiBI gravity is completely equivalent to GR in vacuum, while differs it when the matter is included in. In the presence of matter, this theory shows several attractive new features. In this gravity, the cosmological solutions of the model for homogeneous and isotropic space-times demonstrate that there is a minimum length (and maximum density) at early times, pointing to an alternative theory of the absence of Big Bang singularity in early cosmology \cite{BanadosFerreira}. For positive EiBI coupling parameter $\kappa$, the singularities in gravitational collapse may be prevented due to ``repulsive gravity" effects \cite{Pani}. The theory also supports stable, compact pressureless stars made of perfect fluid. So the parameter $\kappa$ will have a near optimal constraint from the relativistic stars. In Ref. \cite{Delsate}, the authors found that EiBI gravity coupled to a perfect fluid reduces to GR coupled to a nonlinearly modified perfect fluid resulting in an ambiguity between modified coupling and equation of state. Through the energy conditions, consistency, and singularity-avoidance perspectives, they argued that such theory is viable from both an experimental and theoretical point of view.

Although EiBI gravity can resolve the spacetime singularities and is compatible with all current observations, it is still has some shortcomings. For example, it was shown in Ref. \cite{Sotiriou} that it has curvature singularities at the surface of polytropic stars and unacceptable Newtonian limit. The Big Rip singularity was found to be unavoidable in the EiBI phantom model \cite{Bouhmadi-Lopez}. In Refs. \cite{Escamilla} and \cite{Yang}, the transverse and traceless tensor modes were found to be linearly unstable in the perturbations to a four-dimensional homogeneous and isotropic universe. However, in the brane-world scenario, the transverse and traceless tensor fluctuations are stable \cite{Liu}, thus the brane-world scenario may be helpful to stabilize the models in this gravity theory. For the full linear perturbed modes with the approximate background solutions near the maximum density, it was found that \cite{Yang}, for the positive EiBI parameter $\kappa$, the scalar modes are stable in the infinite wavelength limit (the wave vector $k=0$) but unstable for $k\neq0$. The vector modes are stable, while the tensor mode is unstable in the Eddington regime, independent of $k$. However, these modes are unstable and hence cause the instabilities for negative parameter $\kappa$ \cite{Yang}. In Ref. \cite{Lagos}, a general algorithm was proposed and developed to find the action for cosmological perturbations which rivals the conventional, gauge-invariant approach and can be applied to theories with more than one metric. And this method was applied to the the minimal EiBI theory, and the result shows that the tensor-to-scalar ratio of perturbations is unacceptably large. Moreover, many authors have looked at various aspects of this theory \cite{Casanellas,Avelino,Ferreira,ppAvelino,Sham,Cho,Scargill,Jana,Kim,Leung,Harko,Lobo,Sushkov,Olmo,KimKim,Kim3,Lin,Du,Sotani}. And these works show that the new features closely depend on the coupling parameter, especially on the positive $\kappa$.

On the other hand, lensing in strong gravitational field is a powerful tool to test the nature of compact object. Comparing with the lensing in the weak gravitational field case, in the vicinity of a compact object, i.e., a black hole, the lensing will have a rich structure, and a strong field treatment of gravitational lensing was developed in \cite{Bozza,Bozza03prd} based on the lens equation given in Refs. \cite{Frittelli,Virbhadra}. Many study displayed that the property for such lensing is determined by the nature of the compact object. So lensing in strong gravitational field will provide us a possible way to test and distinguish different gravitational theories in the strong gravitational region. The results implied that with the astronomical observation in the near future, we are allowed to verify the alternative theories of gravity in the strong field regime \cite{Eiroa,Sarkar06,ChenJing09,Chen,Kraniotis}. The study has applied to different black hole solutions, and we refer the reader to Ref. \cite{Bozzagrg} and references therein for a recent review.

It is the purpose of this paper to explore the nature of the black hole in EiBI gravity from the viewpoint of the strong gravitational lensing. At first, we construct a static spherically symmetric black hole with a cosmological constant when the electromagnetic field is included in. Then based on this black hole solution with vanishing cosmological constant, we study the lensing in the strong deflection limit by adopting Bozza's method. We find that the presence of the EiBI parameter $\kappa$ with positive value will enlarge the black hole horizon and photon sphere. The strong deflection limit coefficients and observables are obtained.

The paper is structured as follows. In section \ref{Spacetimemetric}, by solving the equations of motion, we obtain a static, spherically symmetric black hole solution with nonvanishing cosmological constant. Next, we investigate the lensing by an asymptotically flat black hole in strong deflection limit in section \ref{stronglensing}. The structure of photon sphere and strong deflection limit coefficients are got. In section \ref{Observables3}, we suppose that the gravitational field of the supermassive black hole at the center of our Galaxy is described by the EiBI black hole and then obtain the numerical results for the main observables in the strong deflection limit. Finally, discussion of our results is undertaken.

\section{Field equations and black hole solution}
\label{Spacetimemetric}

In this section, we generalize the charged black hole solution obtained in Ref.~\cite{BanadosFerreira}.
The action for the EiBI theory is given by
\begin{eqnarray}
 S(g, \Gamma, \Phi)=\frac{1}{8\pi\kappa}\int d^{4}x\bigg(\sqrt{-|g_{\mu\nu}
   +\kappa R_{\mu\nu}(\Gamma)|}-\lambda\sqrt{g}\bigg)+S_{M}(g,\Phi),\label{action}
\end{eqnarray}
where we set $c=G=1$, $R_{\mu\nu}(\Gamma)$ denotes the symmetric part of the Ricci tensor built with the connection $\Gamma$, the dimensionless parameter $\lambda=1+\kappa \Lambda$ corresponds to cosmological constant, and $S_{M}(g,\Phi)$ is the action of matter fields, which only couple to the metric. When the parameter $\kappa\ll g/R$, this action reduces to the Einstein-Hilbert action with cosmological constant $\Lambda$. While when $\kappa\gg g/R$, the Eddington's action will be obtained approximately. Therefore, the EiBI parameter $\kappa$ with inverse dimension of cosmological constant bridges two different gravity theories.

By varying the action (\ref{action}) with respect to the metric field $g$ and connection field $\Gamma$, we get the equations of motion for this gravity theory:
\begin{eqnarray}
 q_{\mu\nu}&=&g_{\mu\nu}+\kappa R_{\mu\nu},\label{auxiliarymetric}\\
 \sqrt{-|q_{\rho\sigma}|}q^{\mu\nu}&=&\lambda\sqrt{-|g_{\rho\sigma}|}g^{\mu\nu}
  -8\pi\kappa\sqrt{-|g_{\rho\sigma}|}T^{\mu\nu},\label{fieldequation}
\end{eqnarray}
where $q_{\mu\nu}$ is the auxiliary metric compatible to the connection with $\Gamma^{\lambda}_{\mu\nu}=\frac{1}{2}g^{\lambda\sigma}(q_{\sigma\mu,\nu}
+q_{\sigma\nu,\mu}-q_{\mu\nu,\sigma})$.

Without the matter fields, the action reduces to the Einstein-Hilbert one. So in order to find the new property of the EIBI theory, we add an electromagnetic field $\mathcal{L}_{M}=-\frac{1}{16\pi}F_{\mu\nu}F^{\mu\nu}$ for simple, for which the energy-momentum is given by
\begin{eqnarray}
 T_{\mu\nu}=\frac{1}{4\pi}(F_{\mu\sigma}F^{\sigma}_{\nu}
        -\frac{1}{4}g_{\mu\nu}F_{\sigma\rho}F^{\sigma\rho}).\label{elect}
\end{eqnarray}
Then the EiBI gravity will deviate from GR. One of the interesting issues is to seek the black hole solution, which will provide a test of different gravity theory in strong gravitational fields.

First, we assume a static spherically symmetric black hole metric,
\begin{eqnarray}
 ds^{2}=-\psi^{2}(r)f(r)dt^{2}+\frac{1}{f(r)}dr^{2}+r^{2}(d\vartheta^{2}+\sin^{2}\vartheta d\phi^{2}).\label{metric}
\end{eqnarray}
In the absent of matter fields, the solution is required to be consistent with the Schwarzschild-de Sitter black hole at large $r$ limit for positive $\Lambda$, i.e.,
\begin{eqnarray}
 \psi(r)&=&1,\label{psir}\\
 f(r)&=&1-2M/r-\Lambda r^{2}/3.\label{fr}
\end{eqnarray}
The auxiliary metric $q_{\mu\nu}$ is assumed as
\begin{eqnarray}
 ds'^{2}=-G^{2}(r)F(r)dt^{2}+\frac{1}{F(r)}dr^{2}+H^{2}(r)(d\vartheta^{2}+\sin^{2}\vartheta d\phi^{2}).
\end{eqnarray}
It corresponds to the spacetime metric (\ref{metric}) with relation (\ref{auxiliarymetric}).
Here we only consider the electrostatic field with nonvanishing $A_{0}$, while other components of the vector field vanish. Then from the Maxwell equation, one easily gets
\begin{eqnarray}
 \partial_{r}(\psi^{-1}r^{2}E)=0,
\end{eqnarray}
where the electric field $E=-\partial_{r}A_{0}$. The solution of this equation is
\begin{eqnarray}
 E=\frac{C_{0}}{r^{2}}\psi(r).
\end{eqnarray}
Here $C_{0}$ is an unfixed constant. The nonvanishing components of the energy-momentum tensor read
\begin{eqnarray}
 8\pi T^{00}&=&\psi^{-4}f^{-1}E^{2},\quad
 8\pi T^{11}=-\psi^{-2}fE^{2},\\
 8\pi T^{22}&=&r^{-2}\psi^{-2}E^{2},\quad
 8\pi T^{33}=r^{-2}\sin^{-2}\theta~\psi^{-2}E^{2}.
\end{eqnarray}
Inserting these into the field equation (\ref{fieldequation}), we have
\begin{eqnarray}
 \frac{H^{2}}{GF}&=&\frac{r^{2}}{\psi f}\bigg(\lambda
               +\frac{\kappa C_{0}^{2}}{r^{4}}\bigg),\\
 H^{2}GF&=&r^{2}\psi f\bigg(\lambda+\frac{\kappa C_{0}^{2}}{r^{4}}\bigg),\\
 G&=&\psi\bigg(\lambda-\frac{\kappa C_{0}^{2}}{r^{4}}\bigg).\label{functionG}
\end{eqnarray}
Solving the first two equations, we obtain
\begin{eqnarray}
 H&=&r\sqrt{\lambda+\frac{\kappa C_{0}^{2}}{r^{4}}},\label{functionH}\\
 F&=&f\bigg(\lambda-\frac{\kappa C_{0}^{2}}{r^{4}}\bigg)^{-1}.\label{functionF}
\end{eqnarray}
It is now clear that the spacetime and auxiliary metrics are related with each other by Eqs. (\ref{functionG})-(\ref{functionF}). Moreover, the field equation (\ref{auxiliarymetric}) reduces to
\begin{eqnarray}
 4\frac{G'}{G}\frac{H'}{H}+2\frac{F'}{F}\frac{H'}{H}+3\frac{G'}{G}\frac{F'}{F}
   +2\frac{G''}{G}+\frac{F''}{F}
     &=&\frac{2}{\kappa F}\bigg(\frac{1}{\lambda-\frac{\kappa C_{0}^{2}}{r^{4}}}-1\bigg),\label{Eq1}\\
 4\frac{H''}{H}+2\frac{F'}{F}\frac{H'}{H}+3\frac{G'}{G}\frac{F'}{F}
   +2\frac{G''}{G}+\frac{F''}{F}
     &=&\frac{2}{\kappa F}\bigg(\frac{1}{\lambda-\frac{\kappa C_{0}^{2}}{r^{4}}}-1\bigg),\label{Eq2}\\
 -\frac{1}{H^{2}F}+\frac{F'}{F}\frac{H'}{H}+\frac{G'}{G}\frac{H'}{H}
                 +\frac{H'^{2}}{H^{2}}+\frac{H''}{H}&=&\frac{1}{\kappa F}
                 \bigg(\frac{1}{\lambda+\frac{\kappa C_{0}^{2}}{r^{4}}}-1\bigg).\label{Eq3}
\end{eqnarray}
From Eqs. (\ref{Eq1}) and (\ref{Eq2}), one has $G'/G=H''/H'$. Thus
\begin{eqnarray}
 G=C_{1}H'.
\end{eqnarray}
Then the metric function $\psi$ and electric field strength are
\begin{eqnarray}
 \psi&=&\frac{C_{1}r^{2}}{\sqrt{\lambda r^{4}+\kappa C_{0}^{2}}}, \label{psi}\\
 E&=&\frac{C_{0}C_{1}}{\sqrt{\lambda r^{4}+\kappa C_{0}^{2}}}. \label{E}
\end{eqnarray}
Requiring $E\rightarrow\frac{Q}{r^{2}}$ when $r\rightarrow \infty$, the constants satisfy
\begin{eqnarray}
 C_{0}C_{1}=\sqrt{\lambda} Q.
\end{eqnarray}
From Eqs.~(\ref{psi}) and (\ref{E}), we have $\psi= \frac{E}{C_{0}}r^2$. When the electric field vanishes, we should have $\psi = 1$. So the constant $C_{0}$ measures the charge of the black hole, and we can set $C_{0}=Q$ and $C_{1}=\sqrt{\lambda}$. Solving Eq. (\ref{Eq3}) yields
\begin{eqnarray}
 F=\frac{1}{HH'^{2}}\bigg[C_{2}+\int\bigg(H'-\frac{H^{2}H'}{\kappa}\bigg(1-\bigg(\lambda+
    \frac{\kappa Q^{2}}{r^{4}}\bigg)^{-1}\bigg)\bigg)dr\bigg].
\end{eqnarray}
Using Eq. (\ref{functionF}), we get the metric function
\begin{eqnarray}
 f(r)&=&\frac{r\sqrt{\lambda r^{4}+\kappa Q^{2}}}
       {\lambda r^{4}-\kappa Q^{2}}
       \bigg[C_{2}+\int\bigg(H'-\frac{H^{2}H'}{\kappa}\bigg(1-\bigg(\lambda+
    \frac{\kappa Q^{2}}{r^{4}}\bigg)^{-1}\bigg)\bigg)dr\bigg]\nonumber\\
    &=&\frac{r\sqrt{\lambda r^{4}+\kappa Q^{2}}}
       {\lambda r^{4}-\kappa Q^{2}}
      \bigg[C_{2}-\int \frac{(\Lambda r^{4}-r^{2}+Q^{2})
      (\lambda r^{4}-\kappa Q^{2})}{r^{4}\sqrt{\lambda r^{4}+\kappa Q^{2}}}dr\bigg].\label{matric2}
\end{eqnarray}
At the large $r$ limit and vanished $Q$, the metric function $f(r)$ reads
\begin{eqnarray}
 f(r)\rightarrow 1+\frac{C_{2}}{\sqrt{\lambda}r}-\frac{\Lambda}{3}r^{2}+\mathcal{O}(r^{-2}).
\end{eqnarray}
Comparing with the expression (\ref{fr}), we can fix the constant $C_2$ as
\begin{eqnarray}
  C_{2}=-2\sqrt{\lambda}M.
\end{eqnarray}

Here we have obtained the solution of the charged black hole in EiBI gravity with nonvanishing $\Lambda$, which is rewritten as follows:
\begin{eqnarray}
 \psi(r) &=& \frac{r^{2}}{\sqrt{r^{4}+(\kappa/\lambda) Q^{2}}}, ~~~~~~~~~\\
 E(r)&=&\frac{Q}{\sqrt{r^{4}+(\kappa/\lambda) Q^{2}}}, \\
 f(r) &=& \frac{r \sqrt{\kappa  Q^2+\lambda  r^4}} {\lambda  r^4-\kappa  Q^2}\bigg[\frac{\left(3 r^2-Q^2-\Lambda r^4\right) \sqrt{\kappa  Q^2+\lambda  r^4}}{3r^3}+\frac{1}{3}\sqrt{\frac{Q^{3}}{\pi\sqrt{\kappa\lambda}}}\Gamma^{2}(1/4)\nonumber\\
  &&+\frac{4}{3} \sqrt{\frac{i Q^3}{\sqrt{\kappa \lambda }}} F(i\text{arcsinh}(\sqrt{\frac{i}{Q}\sqrt{\frac{\lambda}{\kappa}}}r),-1)-2 \sqrt{\lambda} M \bigg].
\end{eqnarray}
The integral in (\ref{matric2}) has been calculated and the constant of integration is added and $F(\Phi,m)=\int_{0}^{\Phi}(1-m\sin^{2}\theta)^{-1/2}d\theta$ with $-\pi/2<\Phi<\pi/2$ represents the first kind elliptic integral. The auxiliary metric can also be obtained from Eqs. (\ref{functionG})-(\ref{functionF}).

When the charge vanishes $Q=0$, we will get the Schwarzschild-AdS black hole solution: $\psi(r)=1$, $f(r)=1-\frac{2M}{r} -\frac{\Lambda}{3}r^{2}$. And taking the coupling parameter $\kappa\rightarrow0$, it will reduces to the RN-AdS/dS solution: $\psi(r)=1$, $E(r)=\frac{Q}{{r^{2}}}$, $f(r)=1-\frac{2M}{r}+ \frac{Q^2}{r^2}-\frac{\Lambda}{3}r^{2}$.

\section{Strong gravitational lensing}
\label{stronglensing}

In this section, we would like to study the effect of the EiBI parameter $\kappa$ on the lensing in strong gravitational field.

\begin{figure}
\centerline{
\includegraphics[width=8cm,height=6cm]{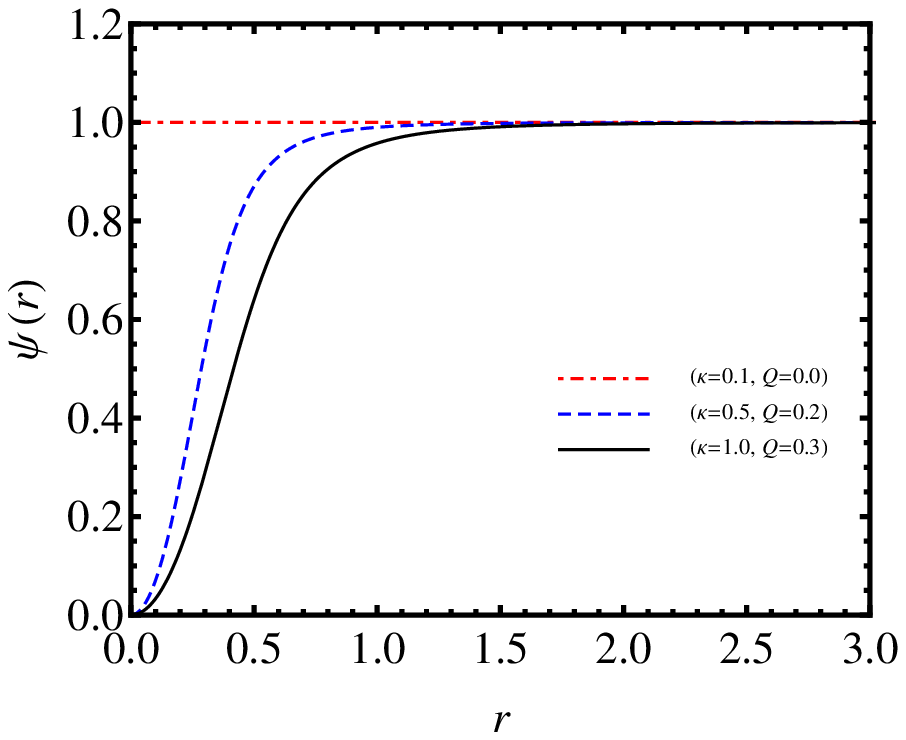}
\includegraphics[width=8cm,height=6cm]{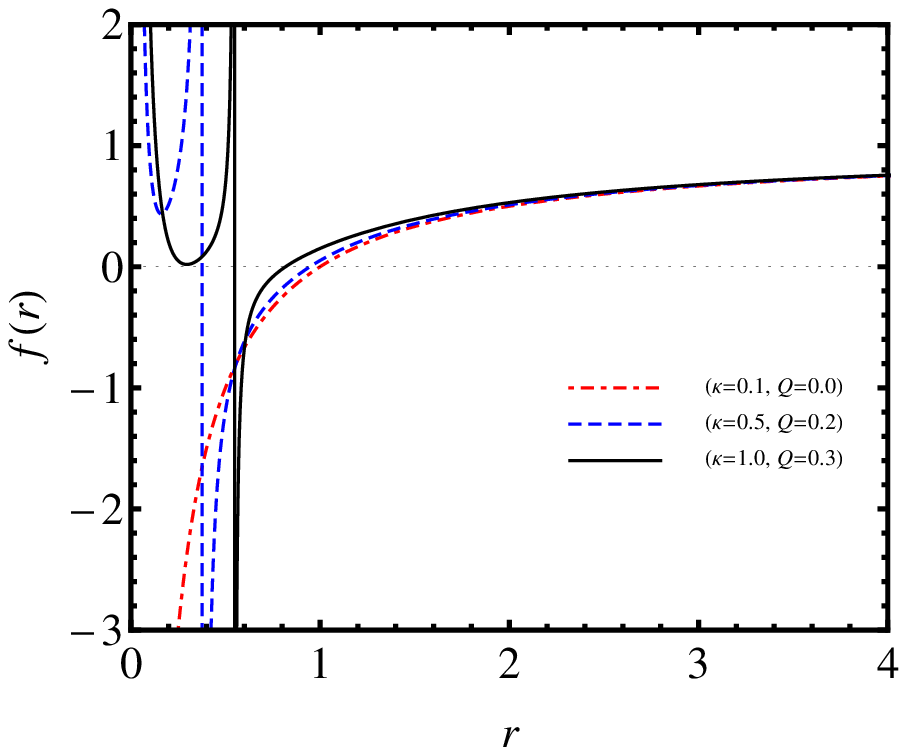}}
\caption{Behaviors of the metric functions $\psi(r)$ and $f(r)$. We have set $M=1/2$ and $\lambda=1$ ($\Lambda=0$).} \label{pmetricfunction}
\end{figure}

\begin{figure}
\centerline{\includegraphics[width=8cm,height=6cm]{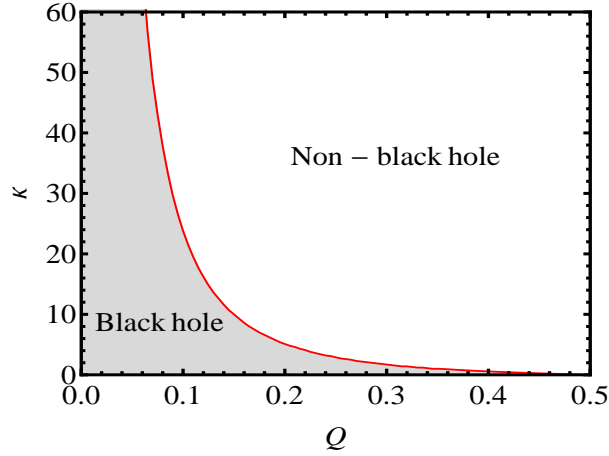}}
\caption{Black hole region in the parameter space. Black hole region locates below the solid line, while above it there exists no black hole.} \label{pbh}
\end{figure}

\begin{figure}
\centerline{\subfigure[]{\label{prh}
\includegraphics[width=8cm,height=6cm]{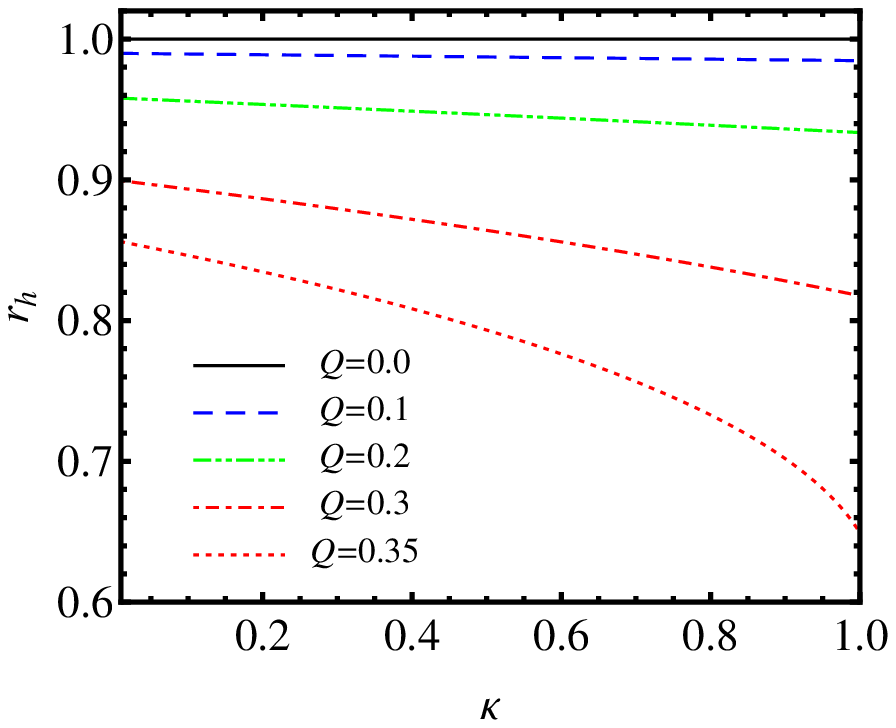}}
\subfigure[]{\label{prps}
\includegraphics[width=8cm,height=6cm]{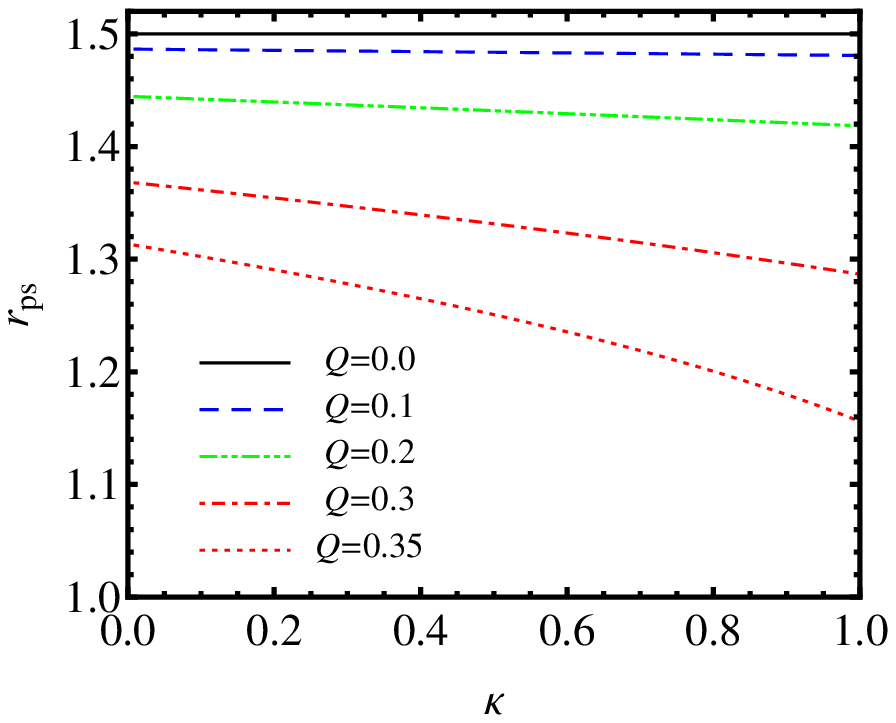}}}
\caption{The radii of the horizon (a) and photon sphere (b) as the functions of the EiBI parameter $\kappa$ for different value of charge $Q$.} \label{pradius}
\end{figure}

Here we only consider an asymptotic flat black hole with $\lambda=1$ ($\Lambda=0$). Then the black hole solution reduces to
\begin{eqnarray}
 \psi(r)&=&\frac{r^{2}}{\sqrt{r^{4}+\kappa Q^{2}}},\label{psi2}\\
  E&=&\frac{Q}{\sqrt{r^{4}+\kappa Q^{2}}},\\
 f(r)&=&\frac{r\sqrt{r^{4}+\kappa Q^{2}}}{r^{4}-\kappa Q^{2}}
   \bigg[\frac{(3r^{2}-Q^{2})\sqrt{r^{4}+\kappa Q^{2}}}{3r^{3}}+\frac{1}{3}\sqrt{\frac{Q^{3}}{\sqrt{\kappa}\pi}}\Gamma^{2}(1/4)\nonumber\\
    &&+\frac{4}{3}\sqrt{\frac{iQ^{3}}{\sqrt{\kappa}}}F(i\text{arcsinh}(\sqrt{\frac{i}{\sqrt{\kappa}Q}}r),-1)
    -2M\bigg].\label{fr2}
\end{eqnarray}
In the limit of $r\rightarrow\infty$, we have
        \begin{eqnarray}
         f(r) &\rightarrow & 1-\frac{2M}{r}+\frac{Q^2}{r^2}+\frac{2\kappa Q^2}{r^4}+\mathcal{O}(r^{-5}).
        \end{eqnarray}
In the limit of $r\rightarrow \sqrt{\sqrt{\kappa}Q}$,
        \begin{eqnarray}
         \psi(r)\rightarrow \frac{1}{\sqrt{2}},~~
         E(r)\rightarrow\frac{1}{\sqrt{2\kappa}},~~
         f(r) \rightarrow \infty.
        \end{eqnarray}
We give three `scalar curvature'
\begin{eqnarray}
 R[g]&\equiv&g^{\mu\nu}R_{\mu\nu}[g] \propto {(r^4-\kappa Q^2)}^{-3},\\
 R[g,q]&\equiv&g^{\mu\nu}R_{\mu\nu}[q]
           =g^{\mu\nu}{(q_{\mu\nu}-g_{\mu\nu})}/\kappa
           ={8/\kappa},\\
 R[q]&\equiv&q^{\mu\nu}R_{\mu\nu}[q]
         =8\frac{(r^4+\kappa Q^2/\sqrt{2})(r^4-\kappa Q^2/\sqrt{2})}
                {(r^4+\kappa Q^2){(r^4-\kappa Q^2)}},
\end{eqnarray}
from which we can see that both $R[g]$ and $R[q]$ are singular at the $r=\sqrt{\sqrt{\kappa}Q}$, but $R[g,q]$ is regular everywhere.
For the small EiBI parameter $\kappa$, we show the behaviors of the metric functions $\psi(r)$ and $f(r)$ for different values of $(\kappa, Q)$ with $M=1/2$ in Fig. \ref{pmetricfunction}. When $Q=0$, we have $\psi(r)=1$ and $f(r)=1-2M/r$, which is exactly consistent with the Schwarzschild black hole as expected. While when the charge $Q$ deviates from zero, the solution will be significantly different from the Schwarzschild one. One remarkable property is that the horizon structure of the charged EiBI black hole is different from the charged black hole in GR, but similar to the Schwarzschild black hole with one horizon. We show the black hole region in the parameter space marked with light gray color in Fig. \ref{pbh}. It clear that black hole with small value of charge can have large $\kappa$. The radius $r_{h}$ of the horizon as a function of the EiBI parameter $\kappa$ is plotted in Fig. \ref{prh}. When the charge $Q$ is small, $r_{h}$ is almost the same for any value of the EiBI parameter $\kappa$. While when $Q$ increases, it displays a monotone decreasing behavior which implies that $\kappa$ shrinks the black hole horizon for fixed charge $Q$. And the horizon radius of the Schwarzschild black hole $r_{h}=1$ will be got with $Q=0$. Moreover, for fixed $\kappa$, the horizon radius decreases with $Q$. Here it is worth noting that the similar asymptotically flat black hole solution was first obtained in Ref.~\cite{BanadosFerreira}, where the metric function is expressed as an integral form.


Adopting the setup $2M=1$, the spacetime metric takes the form
\begin{eqnarray}
 ds^{2}=-A(r)dt^{2}+B(r)dr^{2}+C(r)(d\theta^{2}+\sin^{2}\theta d\phi^{2}),\label{met}
\end{eqnarray}
with the metric functions given by
\begin{eqnarray}
 A(r)=\psi^{2}(r)f(r),\quad
 B(r)=f(r)^{-1}, \quad C(r)=r^{2}.
\end{eqnarray}
Then considering that a photon comes from infinity and returns to infinity, the deflection angle $\alpha(r_{0})$ of the photon for the metric (\ref{met}) was obtained by Virbhadra \emph{et.al.,} \cite{VirbhadraNarasimha}
\begin{eqnarray}
 \alpha(r_{0})&=&I(r_{0})-\pi,\label{alphar0}\\
 I(r_{0})&=&2\int^{\infty}_{r_{0}}
   \frac{\sqrt{B}}{\sqrt{C}\sqrt{\frac{CA_{0}}{C_{0}A}-1}}dr,
\end{eqnarray}
where $r_{0}$ denotes the closest distance that the photon can approach. The subscript ``0'' represents that the metric coefficients are evaluated at $r_{0}$. When the parameter $\kappa$ is fixed, $\alpha(r_{0})$ monotonically decreases with $r_{0}$. When $r_{0}$ approaches some certain points, the photon can complete one loop or more than one loop before reaching the observer located at infinity. When $r_{0}$ approaches the radius $r_{\text{ps}}$ of the photon sphere, the photon will surround the black hole all the time if there is no perturbation.

Virbhadra and Ellis \cite{Virbhadra} and  Claudel \emph{et. al.,} \cite{Claudel} gave two different definitions of photon sphere for a general static spherically symmetric space-time and then they \cite{Claudel}-\cite{Virbhadra22} showed that both definitions give same result $\frac{C'(r)}{C(r)}=\frac{A'(r)}{A(r)}$, where a prime represents a derivative with respect to $r$. For our case, the equation determining the radius $r_{\text{ps}}$ reduces to $A'r=2A$, which takes the form
\begin{eqnarray}
  &&8 \kappa ^{5/4} Q^2 r^2 \left(r^2-Q^2\right) \sqrt{\kappa
   Q^2+r^4} \nonumber\\
   &=&
   \left(3 \kappa ^2 Q^4+2 \kappa  Q^2 r^4+3 r^8\right)\nonumber\\
   &\times&
   \bigg[-4 \sqrt{i} Q^{3/2}rF(i\text{arcsinh}(\sqrt{\frac{i}{\sqrt{\kappa}Q}}r),-1) \nonumber \\
   &-&\frac{Q^{3/2} r \Gamma
   \left(\frac{1}{4}\right)^2}{\sqrt{\pi }}+\sqrt[4]{\kappa}
   \left(3 r-2 \sqrt{\kappa  Q^2+r^4}\right)\bigg]. \label{ps}
\end{eqnarray}
In general, the photon sphere for a static, spherically symmetric black hole is defined as the unstable circular orbit. According to it and after a brief check, the radius of the photon sphere is just the largest solution of Eq. (\ref{ps}). When the charge $Q$ vanishes, the solution reduces to $r_{\text{ps}}=3/2$, which is consistent with that of the Schwarzschild black hole. While for a nonvanishing $Q$, the equation cannot be analytically solved. We show the numerical results in Fig. \ref{prps}. The radius $r_{\text{ps}}$ of the photon sphere shares the similar behavior as $r_{h}$, while has a larger value than it. From the figure, it is clear that both $Q$ and $\kappa$ shrinks the photon sphere. Thus the photon is more easily captured by the black hole with small $Q$ and $\kappa$.

\begin{figure*}
\subfigure[]{\label{pa1}
\includegraphics[width=8cm,height=6cm]{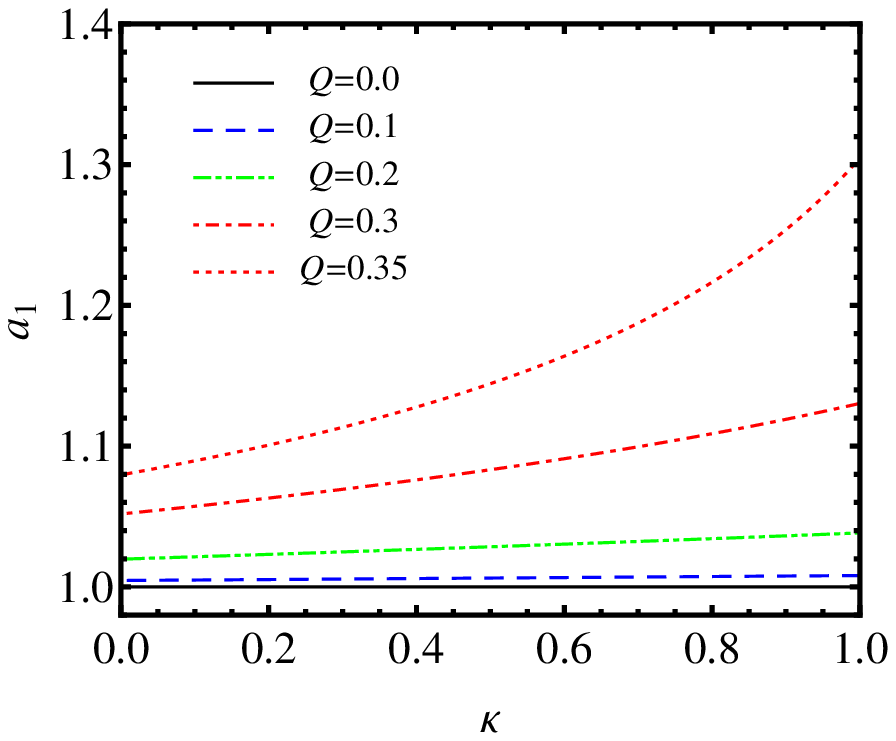}}
\subfigure[]{\label{pa2}
\includegraphics[width=8cm,height=6cm]{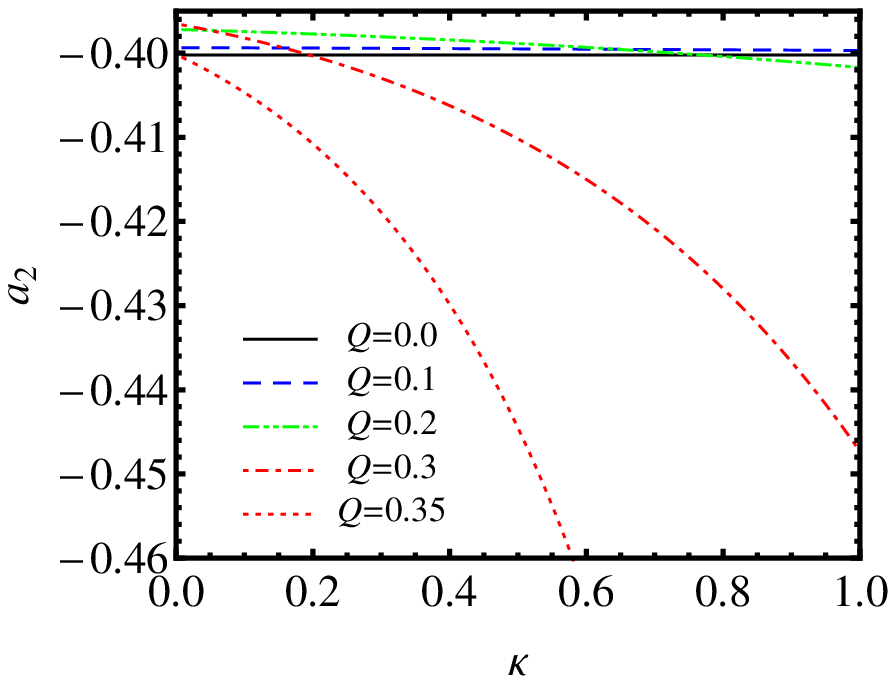}}\\
\centerline{\subfigure[]{\label{pups}
\includegraphics[width=8cm,height=6cm]{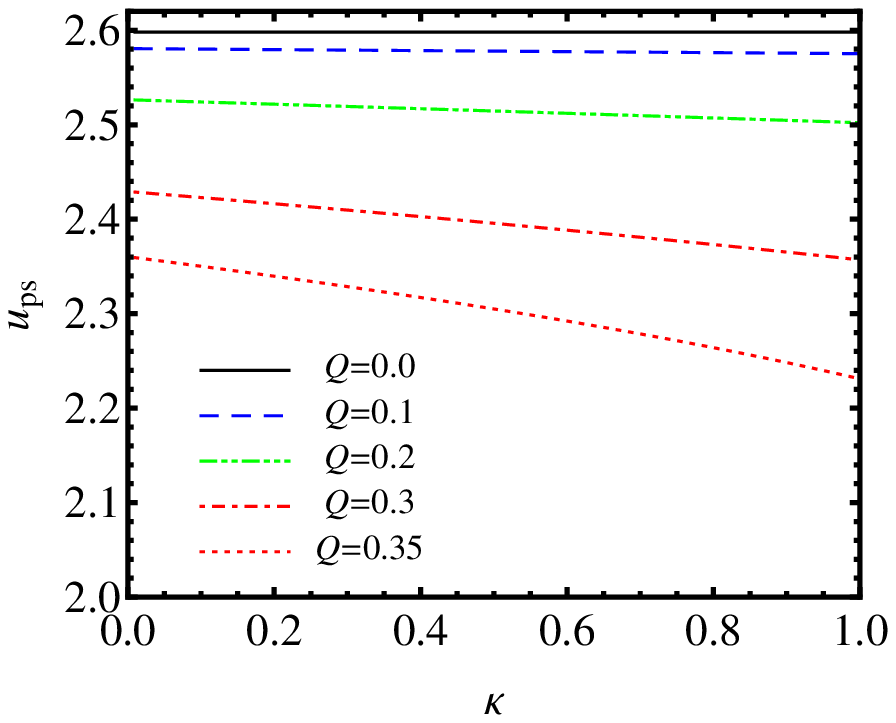}}}
\caption{Variation of the strong deflection limit coefficients and minimum impact parameter $u_{\text{ps}}$. (a) $a_{1}$,  (b) $a_{1}$, and (c) $u_{ps}$.} \label{Sdlc}
\end{figure*}

Next, let us consider the lensing that the photon passes very close the photon sphere. Following the method proposed by Bozza \cite{Bozza}, we define a variable
\begin{eqnarray}
 z=1-\frac{r_{0}}{r},
\end{eqnarray}
then the total azimuthal angle is expressed as
\begin{eqnarray}
 I(r_{0})=\int^{1}_{0}R(z,r_{0})K(z,r_{0})dz,\label{Ir0}
\end{eqnarray}
where
\begin{eqnarray}
 R(z,r_{0})&=&\frac{2r_{0}\sqrt{ABC_{0}}}{C(1-z^{2})},\\
 K(z,r_{0})&=&\frac{1}{\sqrt{A_{0}-AC_{0}/C}}.\label{K0}
\end{eqnarray}
After a simple check, we find that $R(z,r_{0})$ is a regular function of $z$ and $r_{0}$, while $K(z,r_{0})$ diverges at $z=0$. To deal with this problem, we split (\ref{Ir0}) into two parts, the divergent part $I_{\text{D}}(r_{0})$ and regular part $I_{\text{R}}(r_{0})$,
\begin{eqnarray}
 I_{\text{D}}(r_{0})&=&\int_{0}^{1} R(0,r_{\text{ps}})K_{0}(z,r_{0})dz,\\
 I_{\text{R}}(r_{0})&=&\int^{1}_{0}\bigg(R(z,r_{0})K(z,r_{0})-R(0,r_{\text{ps}})K_{0}(z,r_{0})\bigg)dz.
\end{eqnarray}
In order to find the order of divergence of the integrand, we perform a Taylor expansion of the argument inside the square root in $K(z,r_{0})$,
\begin{eqnarray}
 K_{0}(z,r_{0})=\frac{1}{\sqrt{\chi_{1}(r_{0})z+\chi_{2}(r_{0})z^{2}+\mathcal{O}(z^{3})}},
\end{eqnarray}
where the coefficients $\chi_{1}(r_{0})$ and $\chi_{2}(r_{0})$ cannot be shown in a compact form, and so we do not list them. However, one thing worth to note is that the coefficient $\chi_{1}(r_{0})$ vanishes at $r_{0}=r_{\text{ps}}$, which leads to a logarithmic divergence of the deflection angle (\ref{alphar0}). Thus, the deflection angle can be expanded with a logarithmic term presented by Bozza \cite{Bozza},
\begin{eqnarray}
 \alpha(u)=-a_{1}\log\bigg(\frac{u}{u_{\text{ps}}}-1\bigg)
                +a_{2}+\mathcal{O}(u-u_{\text{ps}}).\label{alphau}
\end{eqnarray}
The strong deflection limit coefficients $a_{1}$ and $a_{2}$ depend on the radius of the photon sphere,
\begin{eqnarray}
 a_{1}&=&\frac{R(0,r_{\text{ps}})}{2\sqrt{\chi_{2}(r_{\text{ps}})}},\\
 a_{2}&=&-\pi+a_{\text{R}}+a_{1}\log\frac{r_{\text{ps}}^{2}(2A(r_{\text{ps}})-r_{\text{ps}}^{2}A''(r_{\text{ps}}))}
         {u_{\text{ps}}r_{\text{ps}}A^{3/2}(r_{\text{ps}})},\\
 a_{\text{R}}&=&I_{\text{R}}(r_{\text{ps}}).
\end{eqnarray}
In general, the coefficient $a_{\text{R}}$ can not be calculated analytically, but we can get it numerically. The minimum impact parameter is in the form
\begin{eqnarray}
 u_{\text{ps}}=\frac{r_{\text{ps}}}{\sqrt{A(r_{\text{ps}})}}.
\end{eqnarray}
The coefficients $a_{1}$, $a_{2}$, and $u_{\text{ps}}$ are plotted in Fig. \ref{Sdlc}. When the charge $Q=0$ (described by the black solid lines), we have $a^{\text{Sch}}_{1}=1$, $a^{\text{Sch}}_{2}=-0.4002$, and $u^{\text{Sch}}_{\text{ps}}=3\sqrt{3}/2$ for the Schwarzschild black hole. When the nonvanishing charge $Q$ is fixed, it is shown that the coefficient $a_{1}$ increases, while $u_{\text{ps}}$ decreases with the EiBI parameter $\kappa$. For fixed $\kappa$, $a_{1}$ increase while $u_{\text{ps}}$ decrease with the charge $Q$. The variation of the strong field limit coefficients $a_{2}$ is also a monotonic function of $\kappa$. It decreases quickly for larger $Q$.

\begin{figure}
\centerline{\includegraphics[width=8cm,height=6cm]{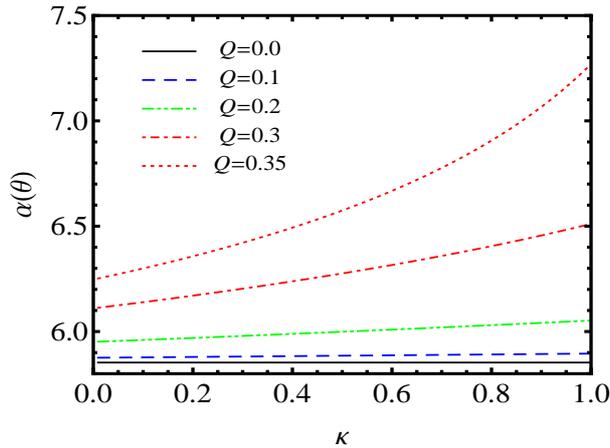}}
\caption{The deflection angle as a function of $\kappa$ for fixed $u=u_{\text{ps}}+0.005$.} \label{palpha}
\end{figure}

Figure \ref{palpha} shows the deflection angle $\alpha(u)$ evaluated at $u=u_{\text{ps}}+0.005$. It indicates that $\kappa$ increases $\alpha(u)$ for the light propagated in the charged EiBI black hole background. And charge $Q$ will further increase $\alpha(u)$. It also implies that the presence $\kappa$ and $Q$ will give a larger $\alpha(u)$ than that of the Schwarzschild black hole, i.e., $\alpha(u)=5.85$. Therefore, comparing with those in the Schwarzschild black hole, we could extract the information about the size of the EiBI parameter $\kappa$ by using the strong gravitational lensing.

On the other hand, the lens geometry is a key ingredient to investigate the black hole lensing, i.e., the positions and magnifications of the relativistic images. So it is necessary to give a brief introduction to the lens geometry. The lens configuration is assumed that the black hole lens is located between the light source and observer, and both of them are required to be far from the black hole, so that the gravitational fields there are very weak and the spacetime can be described by a flat metric. Optical axis for such lens configuration is defined as the line joining the observer and the lens. We denote the angular positions of the source (S) and the images (I), seen from the observer, as $\varpi$ and $\theta$. Then the lens equation is shown in the form \cite{Virbhadra}:
\begin{eqnarray}
 \tan\varpi=\tan\theta-\frac{D_{\text{LS}}}{D_{\text{OS}}}\big[\tan(\alpha-\theta)
                    +\tan\theta\big].\label{lensgeometry}
\end{eqnarray}
Here $D_{\text{OL}}$, $D_{\text{LS}}$ and $D_{\text{OS}}$ correspond to the observer-lens, lens-source, and observer-source distances, respectively.

Considering an idealized model that the source, lens, and observer are highly aligned, then the lens equation is simplified as
\begin{eqnarray}
 \varpi=\theta-\frac{D_{\text{LS}}}{D_{\text{OS}}}\Delta\alpha_{n}.
 \label{varpi}
\end{eqnarray}
Based on the lens geometry, one gets $u=D_{\text{OL}}\sin\theta\sim D_{\text{OL}}\theta$. Thus the deflection angle (\ref{alphau}) can be reexpressed as
\begin{eqnarray}
 \alpha(\theta)=-a_{1}\log\big(\frac{\theta D_{\text{OL}}}{u_{\text{ps}}}-1\big)
                +a_{2}.\label{A2}
\end{eqnarray}
Expanding (\ref{A2}) around $\alpha=2n\pi$ and solving $\theta$, we obtain the position of the $n$-th relativistic images
\begin{eqnarray}
 \theta_{n}=\frac{u_{\text{ps}}}{D_{\text{OL}}}(1+e_{n})
      -\frac{u_{\text{ps}}e_{n}}{a_{2}D_{\text{OL}}}\Delta\alpha_{n},
\end{eqnarray}
where $e_{n}=e^{\frac{{a_2}-2n\pi}{{a_1}}}$. After substituting Eq. (\ref{varpi}) into the above equation, the expression of the position reduces to
\begin{eqnarray}
 \theta_{n}=\theta_{n}^{0}+\frac{u_{\text{ps}}e_{n}}
         {a_{1}}\frac{D_{\text{OS}}}{D_{\text{OL}}D_{\text{LS}}}
            (\varpi-\theta_{n}^{0}),\label{thetan}
\end{eqnarray}
where $\frac{u_{\text{ps}}}{D_{\text{OL}}}\ll 1$ is considered, and $\theta_{n}^{0}=\frac{u_{ps}}{D_{OL}}(1+e_{n})$.

Besides the angular position, the magnification is also an important quantity to describe the relativistic images, which is defined as
\begin{eqnarray}
 \mu_{n}=\left|\frac{\varpi}{\theta_{n}}\frac{d\varpi}{d\theta_{n}}\right| ^{-1}.
\end{eqnarray}
After considering $\frac{u_{\text{ps}}}{D_{\text{OL}}}\ll 1$, we get
\begin{eqnarray}
 \mu_{n}=\frac{e_{n}(1+e_{n})}{{a_1}\varpi}
         \frac{D_{\text{OS}}}{D_{\text{OL}}} \bigg(\frac{u_{\text{ps}}}{D_{\text{OL}}}\bigg)^{2}.\label{mun}
\end{eqnarray}
It is clear that $\mu_{n}\sim e^{-n}$, so the first relativistic image is the brightest one. However, $\mu_{n}\sim (\frac{u_{\text{ps}}}{D_{\text{OL}}}\big)^{2}$, which implies the images are faint.

\section{Observational gravitational lensing parameters}
\label{Observables3}

The observables must be constructed in order to compare with astronomical observations. Here we follow Ref. \cite{Bozza} with the suppose that the outermost relativistic image $\theta_{1}$ is resolved as a single image and the remaining ones are packed together at $\theta_{\infty}$. Then there are three observables, $\theta_{\infty}$ denotes the positions of the relativistic images except the first one, $s$ corresponds to the angular separation between the first image and other ones, and $\tilde{r}$ is the ratio between the flux of the first image and the sum of the others. They are shown in the following forms
\begin{eqnarray}
 \theta_{\infty}&=&\frac{u_{\text{ps}}}{D_{\text{OL}}},\label{thetainfty}\\
 s&=&\theta_{\infty}e^{\frac{{a_2}-2\pi}{{a_1}}}, \label{ss}\\
 \tilde{r}&=&e^{2\pi/{a_1}}. \label{rr}
\end{eqnarray}
For a given theoretical model, the strong deflection limit coefficients $a_{1}$ and $a_{2}$, and the minimum impact parameter $u_{\text{ps}}$ can be obtained. Then these three observables can be calculated. On the other hand, comparing them with astronomical observations, it will allow us to determine the nature of the black hole stored in the lensing.

Next, for an example, we would like to numerically estimate the values of these observables of the gravitational lensing in the strong deflection limit. The lens is supposed to be the supermassive black hole located at the center of our Milk Way, and it is described by the EiBI black hole metric (\ref{metric}) with the metric functions given in (\ref{psi2})-(\ref{fr2}). The mass of the black hole is estimated to be $M=4.4\times 10^{6}M_{\odot}$ with $M_{\odot}$ the mass of the sun and its distance from us is around $D_{\text{OL}}=8.5$ kpc \cite{Genzel}. Then we can estimate the values of the coefficients and observables in strong gravitational
lensing in the EiBI black hole spacetime.

\begin{table}[h]
\begin{center}
\begin{tabular}{cccccccc}
  \hline\hline
      $\kappa$ & $Q$ & $\theta_{\infty}$($\mu arcsecs$) & $s$($\mu arcsecs$) & $r_{m} (magnitudes)$  & $u_{\text{ps}}/R_{s}$ & $a_{1}$ & $a_{2}$ \\
\hline
 Sch-BH& 0.0 & 26.510 & 0.0332 & 6.8219 & 2.598 & 1.000 & -0.4002 \\\hline
    & 0.1 & 26.311 & 0.0340 & 6.7909 & 2.581 & 1.005 & -0.3993 \\
    & 0.2 & 25.779 & 0.0368 & 6.9899 & 2.526 & 1.020 & -0.3972 \\
RN-BH& 0.3 & 24.788 & 0.0433 & 6.4858 & 2.429 & 1.052 & -0.3965\\
    & 0.35 & 24.084 & 0.0493 & 6.3190 & 2.360 & 1.080 & -0.4001 \\\hline
    & 0.1 & 26.326 & 0.0341 & 6.7886 & 2.580 & 1.004 & -0.3994 \\
    & 0.2 & 25.751 & 0.0372 & 6.6789 & 2.524 & 1.021 & -0.3974 \\
 0.1& 0.3 & 24.722 & 0.0445 & 6.4522 & 2.423 & 1.057 & -0.3982\\
   & 0.35 & 23.980 & 0.0518 & 6.2609 & 2.350 & 1.090 & -0.4068 \\\hline
    & 0.1 & 26.304 & 0.0344 & 6.7791 & 2.578 & 1.006 & -0.3995 \\
    & 0.2 & 25.658 & 0.0387 & 6.6327 & 2.515 & 1.029 & -0.3989 \\
 0.5& 0.3 & 24.445 & 0.0507 & 6.2975 & 2.396 & 1.083 & -0.4102\\
   & 0.35 & 23.519 & 0.0658 & 5.9611 & 2.305 & 1.144 & -0.4445 \\\hline
    & 0.1 & 26.277 & 0.0347 & 6.7669 & 2.575 & 1.008 & -0.3997 \\
    & 0.2 & 25.531 & 0.0408 & 6.5698 & 2.502 & 1.038 & -0.4017 \\
 1.0& 0.3 & 24.049 & 0.0624 & 6.0350 & 2.357 & 1.130 & -0.4473\\
   & 0.35 & 22.764 & 0.1087 & 5.2286 & 2.231 & 1.305 & -0.6902 \\\hline\hline
\end{tabular}
\caption{Numerical estimation for the observables and the strong deflection limit coefficients for the Schwarzschild, RN black holes, and the charged EiBI black holes supposed to describe the object at the center of our Galaxy. $r_{m}=2.5\log\tilde{r}$.}\label{tab1}
\end{center}
\end{table}

\begin{figure}
\subfigure[]{\label{Ptheta}
\includegraphics[width=8cm,height=6cm]{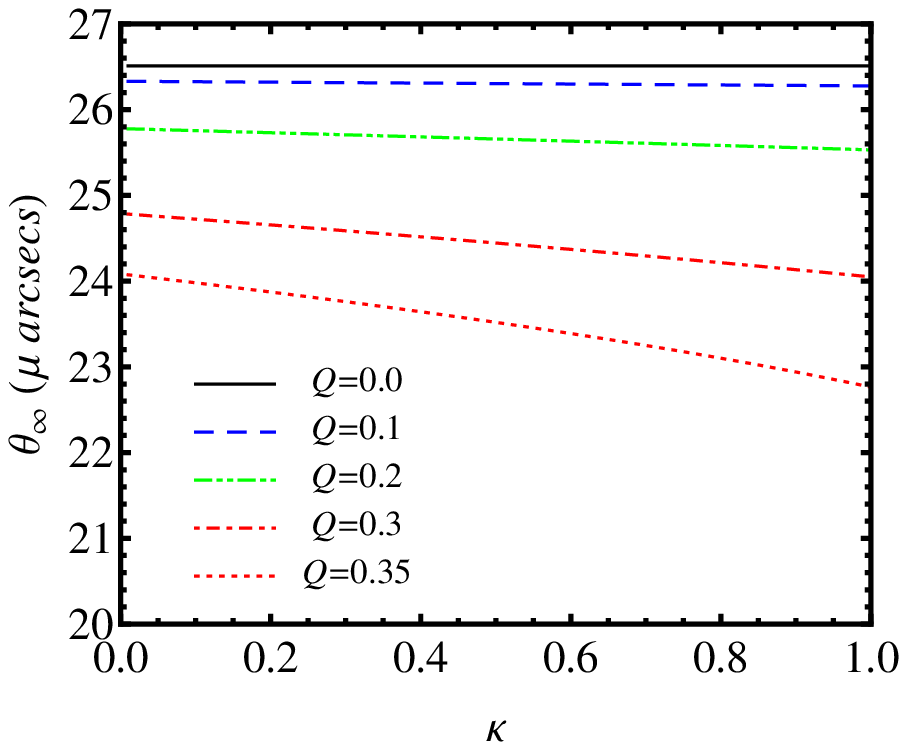}}
\subfigure[]{\label{Ps}
\includegraphics[width=8cm,height=6cm]{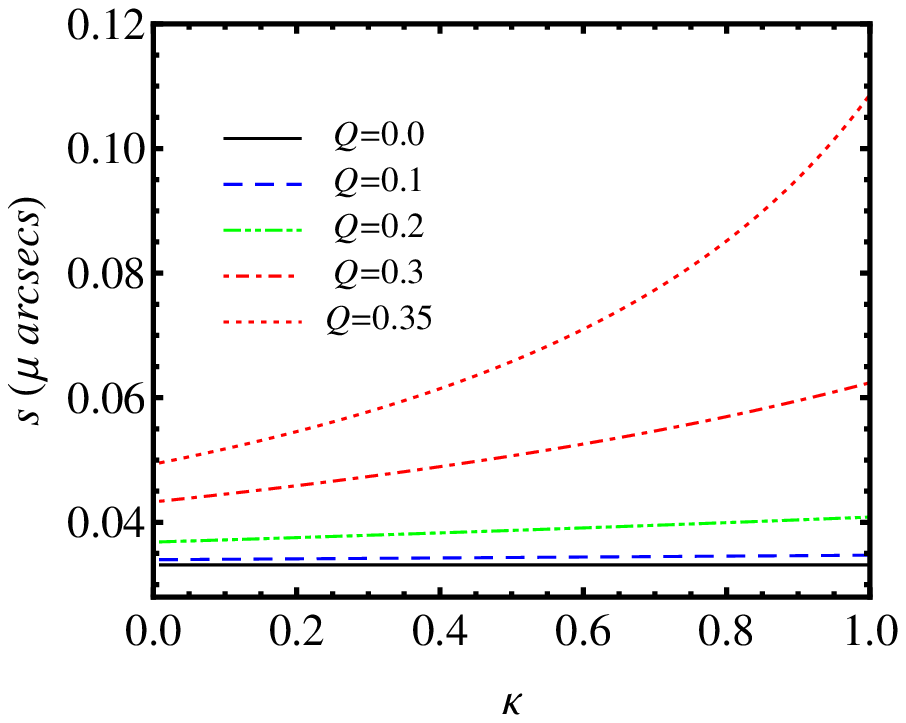}}\\
\centerline{\subfigure[]{\label{Prm}
\includegraphics[width=8cm,height=6cm]{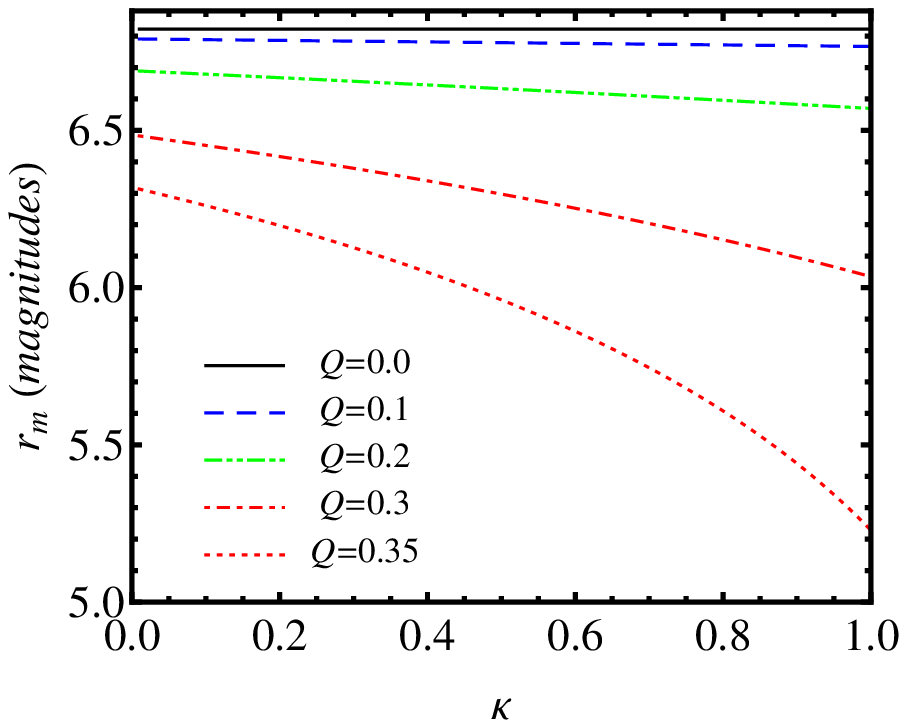}}}
\caption{Variation of the angular position of the relativistic images $\theta_{\infty}$ (a), the relative magnitudes $r_{m}$ (b), and the angular separation $s$ (c) with the EiBI parameter $\kappa$ for fixed charge $Q$.} \label{Pobservables}
\end{figure}

In table \ref{tab1}, we list the numerical values of the observables and the strong field limit coefficients for the Schwarzschild, charged RN, and EiBI black holes. Moreover, the behaviors of the observables are shown in Fig. \ref{Pobservables}. From it, we find that with the increase of $\kappa$, the angular position of the relativistic images $\theta_{\infty}$ and relative magnitudes $r_{m}$ decrease. However, the angular separation $s$ increases with $\kappa$ for the fixed nonvanishing charge $Q$. For a fixed value of $\kappa$, $\theta_{\infty}$ and $r_{m}$ decrease with the charge $Q$, while $s$ increases with it. Moreover, from table \ref{tab1}, we find that, for the fixed charge $Q$, the relations $\theta_{\infty}^{\text{EiBI}}<\theta_{\infty}^{\text{RN}}<\theta_{\infty}^{\text{Sch}}$, $r_{m}^{\text{EiBI}}<r_{m}^{\text{RN}}<r_{m}^{\text{Sch}}$, and $s^{\text{Sch}}<s^{\text{RN}}<s^{\text{EiBI}}$ hold. Compared with the RN and Schwarzschild black holes, the charged EiBI black hole has small $\theta_{\infty}$ and $r_{m}$, while has large $s$. It is illustrated in the table  \ref{tab1} that the observable $\theta_{\infty}$ of RN and Schwarzschild black holes is of $2$ $\mu$arcsecs, and the EiBI black hole is of $4$ $\mu$arcsecs from the Schwarzschild one. Although such difference could further increase with $Q$ and $\kappa$, it is still too small to be tested with modern astronomical instruments \cite{Eckart,BozzaMancini}. Thus distinguishing the EiBI black hole from a GR one may be a challenging task for the next generation of astronomical instruments in the near future.

\begin{figure}
\centerline{\includegraphics[width=8cm,height=6cm]{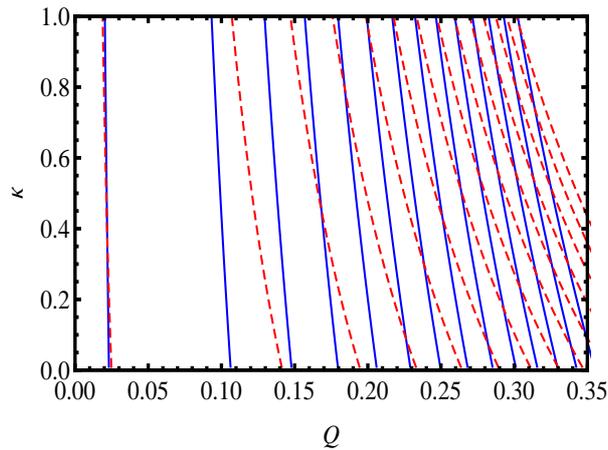}}
\caption{Contour plots of the observables $\theta_{\infty}$ and $r_{m}$ in the plane $(Q, \kappa)$. $\theta_{\infty}$ is described by the blue solid lines with values $24-26.5$ $\mu arcsecs$ from left to right. $r_{m}$ is described by the red dashed lines with values $6.02-6.82$ $magnitudes$ from right to left.} \label{pqkappa}
\end{figure}

As we know, it will be much accurate to determine the nature of the black hole through the lensing using more than one observable. Since the observables $\theta_{\infty}$ and $r_{m}$ are easily observed, we will introduce a possible way to do this by using them. In Fig. \ref{pqkappa}, we plot the contour curves of constant $\theta_{\infty}$ and $r_{m}$ in the plane $(Q, \kappa)$. So each point in the figure is characterized by four values, i.e., $Q$, $\kappa$, $\theta_{\infty}$, and $r_{m}$. By fixing $\theta_{\infty}$ and $r_{m}$ from observations, we are allowed to accurately read the charge $Q$ and EiBI parameter $\kappa$ of the black hole through Fig. \ref{pqkappa}, where the contour curves of constant $\theta_{\infty}$ and $r_{m}$ intersect. Therefore this method could give an accurate test of the parameters of the black hole with combining two observables. Form the figure, we see that large $\kappa$ and $Q$ can be more accurate determined with $\theta_{\infty}$ and $r_{m}$.

\section{Conclusions}
\label{Conclusions}

In this paper, we first obtained a charged EiBI black hole solution with a nonvanishing cosmological constant when the electromagnetic field is included in. The property of such black hole is qualitatively different from the charged RN black hole in GR. The EiBI parameter $\kappa$ was found to shrink the black hole horizon and photon sphere with a monotone decreasing behavior. When the matter fields vanish, such solution reduces to the Schwarzschild black hole (see the black solid lines in Fig. \ref{pradius}).

Based on the solution, we explored the lensing by a asymptotically flat black hole in the strong deflection limit. The result shows that the strong deflection limit coefficient $a_{1}$ increases, while $a_{2}$ and $u_{\text{ps}}$ decreases with the EiBI parameter $\kappa$ for fixed charge $Q$. For a fixed impact parameter $u_{\text{ps}}$, the deflection angle increases with $\kappa$. Following Ref.~\cite{Bozza}, we got three observables, $\theta_{\infty}$, $s$, and $\tilde{r}_{m}$, which measure different properties of the relativistic images. With the increase of $\kappa$, $\theta_{\infty}$ and $\tilde{r}_{m}$ monotonously decrease, while $s$ increases in the EiBI black hole spacetime. With the suppose that the supermassive black hole at our galaxy is described by the EiBI black hole, we found that the observable $\theta_{\infty}$ measuring the positions of the relativistic images is smaller than that of the Schwarzschild and RN black hole. For large $\kappa$, the observables has a large differences between the EiBI black hole and GR one. Therefore these results, in principle, may provide a possibility to test how an astronomical EiBI black hole deviates from a GR black hole in the future astronomical observations.

\section*{Acknowledgement}

This work was supported by the National Natural Science Foundation of China (Grants No. 11205074 and No. 11375075).

\end{document}